\documentclass{elsart}

\usepackage{epsfig}

\begin{document}

\begin{frontmatter}

\title{Phenomena of spin rotation and oscillation of particles (atoms, molecules)
containing in a trap blowing on by wind of high energy particles in storage ring.}

\author{V. Baryshevsky}

\address{Institute of Nuclear Problems, Belarusian State University \\
St. Bobryiskaya 11, 220050, Minsk, Republic of Belarus }

\begin{abstract}
Spin rotation and oscillation phenomena of particles captured in a gas target 
through which beam of 
high energy particles passes is discussed. Such experiment arrangement make it 
realizable for storage ring and allows to study zero-angle scattering amplitude
 at highest possible energies. 
\end{abstract}


\end{frontmatter}


Investigation of spin-dependent interactions of elementary particles
at high energies is a very important part of program of scientific research
has been preparing for carry out at storage rings (RHIC, CERN, COSY). 
There are several experimental possibilities for the indirect measurement of the real part 
of the forward scattering amplitude \cite{Lehar}. Since no scattering experiment is possible in the forward direction, the determination of the 
real part of the forward amplitudes has always consisted in the measurement of well chosen elastic 
scattering observables at small angles and then in the extrapolation of these observables towards 
zero angle \cite{Lehar}.

It has been shown in [2-4]
that there is an unambiguous method which makes the direct measurement
of the real part of the spin-dependent forward scattering amplitude.
This technique is based on the effect of proton (deuteron) beam spin rotation 
in a polarized (nonpolarized) 
nuclear target.
This technique uses the measurement of angle of spin rotation of high energy proton 
(neutron) in
conditions of transmission experiment - the so-called spin rotation experiment .

In the present paper it is shown that we can reverse experiment arrangement to study 
spin rotation and oscillation of particles of gas target through which beam of 
high energy particles passes. Such experiment arrangement make it realizable for 
storage ring and allows to study zero-angle scattering amplitude at highest possible 
energies. 


As a result of numerous studies (see, for example, \cite{12,Goldberger}), a
close connection between the coherent elastic scattering amplitude $f(0)$
and the refraction index of a medium $N$ has been established:
\[
N^2=1+\frac{4\pi \rho }{k^{2}}f\left( 0\right)   
\label{refr_ind2}
\]
where $\rho $ is the number of particles per $cm^{3}$, $k$ is the wave
number of a particle incident on a target.

Let us consider refraction on the vacuum-medium boundary. Wave number of a particle in vacuum is denoted $k$, 
$k^{\prime} = k N$ is the wave number of particle in medium.
The kinetic energy of a particle in vacuum 
$E = \frac{\hbar^2 k^2}{2 m}$ is not equal to that in medium
$E^{\prime} = \frac{\hbar^2 k^2 N^2}{2 m}$. We immediately obtain from the energy conservation 
condition the necessity to suppose that a particle  
in a medium possesses effective potential energy. 
This energy can be found easily from 
evident equality
\begin{equation}
E=E^{\prime}+U ~ \hbox{i.e.}~ U=E-E^{\prime }=- \frac{2 \pi {\hbar}^2}{m} {\rho} f(0).
\label{U-}
\end{equation}
In 1964 it was shown \cite{1rot}
that while slow neutrons are propagating through the target with polarized
nuclei a new effect of nucleon spin precession occurred. It is stipulated by
the fact that neutrons in a polarized target possesses two
refraction indices: ($N_{\uparrow \uparrow }$ for neutrons with the spin
parallel to the target polarization vector and $N_{\uparrow \downarrow }$
for neutrons with the opposite spin orientation , $N_{\uparrow \uparrow
}\neq N_{\uparrow \downarrow })$. 

From (\ref{U-}) it follows that effective potential energy of particle 
$U_{\uparrow \uparrow}~(U_{\uparrow \downarrow }) $
with 
spin parallel (antiparallel) to $\overrightarrow{p}$ in polarized medium is
\begin{equation}
U_{\uparrow \uparrow }=-\frac{2 \pi \hbar^2}{m} \rho f_{\uparrow \uparrow},~~
U_{\uparrow \downarrow }=-\frac{2 \pi \hbar^2}{m} \rho f_{\uparrow \downarrow}.
\end{equation}
The frequency of nucleon spin precession around polarization vector is
$\Omega=\frac{Re (U_{\uparrow \uparrow}-U_{\uparrow \downarrow})}{\hbar}$ and
the angle of rotation can be expressed as
\begin{equation}
\vartheta =\Omega T=-\frac{2\pi \hbar}{m} \rho {Re}\left( f_{\uparrow \uparrow
}-f_{\uparrow \downarrow }\right) T  \label{theta}
\end{equation}
where $T$ is the time of particle being in polarized media.

According to \cite{1rot}, neutron spin precession implies that in the
target with polarized nuclei a nuclear pseudomagnetic field exists and the
interaction of an incident neutron with this field results in neutron spin
rotation. The results obtained in \cite{1rot}, initiated experiments which
proved the existence of this effect [8-10].

Suppose that gas cell contains nucleons and the beam of polarized proton (deuterons) 
passes through it. In this case particle beam is considered as moving polarized target and $T$ makes
sense of time of beam action on an atom in trap. Validity of this statement can be 
easy verified in beam coordinate frame, in which we have routine arrangement - target 
rests while particles captured in a trap move in target.
Therefore we can conclude that spin of nucleons (atoms) captured in a cell rotates in polarized medium 
formed by particle beam moving in a storage ring and angle of rotation is determined by (\ref{theta}).

%
Potential energy of interaction of deuteron with a particle beam passing 
through a gas trap and external magnetic 
field $\overrightarrow{B}$ can be expressed as follows:
\begin{equation}
\hat{U}=-\frac{2\pi {\hbar}^2 }{m} \rho \hat{f}\left( 0\right)-\overrightarrow{{\mu}_d}\cdot \overrightarrow{B}
\label{U}
\end{equation}
where $\overrightarrow{{\mu}_d}$ is the operator of deuteron magnetic moment.
It should be mentioned that if deuteron in a trap exists as deuteron atom then
(\ref{U}) can be written as
\begin{equation}
\hat{U}=-\frac{2\pi {\hbar}^2 }{m} \rho {\hat{f}}_d \left( 0\right)-
\overrightarrow{{\mu}_d}\cdot \overrightarrow{B}-\overrightarrow{{\mu}_e}\cdot \overrightarrow{B}
\label{U2}
\end{equation}
where ${\hat{f}}_d$ is the scattering amplitude of high energy particle or deuteron atom 
(it depends both on spin of high energy particle (deuteron) spin and spin of atom electron).
In this case, to find the angle of atom spin rotation, we should consider Hamiltonian 
${\hat{H}}_A={\hat{H}}_d+{\hat{U}}$, where ${\hat{H}}_d$ is the spin Hamiltonian, describing 
hyperfine interaction in deuteron atom.

The explicit expression for the amplitude $\hat{f}\left( 0\right) $
for particles with arbitrary spin $S$ \ has been obtained in \cite{6rot}.
Using this amplitude we can explicitly express $\hat{U}$ and find the behavior of
spin wave function and other spin characteristics of a particle (atom)in 
any given time moment with the help of Shr\"{o}dinger equation. 
\begin{equation}
i \hbar \frac{\partial \psi }{\partial t}=({\hat{H}}_A+{\hat{U}}) \psi.
\label{Shr_eq}
\end{equation}
where ${\hat{H}}_d$ is the spin Hamiltonian of atom in the ground state, describing 
hyperfine interaction of electron and nucleus in deuteron atom.
Let us consider the simplest case when beam of nonpolarized particles
passes through a trap containing deuterons in the absence of external magnetic field. 
In this case $\hat{f}(0)$ is a function of
deuteron spin operator $\hat{S}$ and can be written as
\begin{equation}
\hat{f}\left( 0\right)
=d+d_{1}S_{z}^{2}  \label{f_hat}
\end{equation}
The quantization axis $z$ is directed along $\vec{n}=\frac{\overrightarrow{k}%
}{k}$, $\overrightarrow{k}$ is the relative momentum of particles. 
Consider a specific case of strong interactions invariant under space
and time reflections. For this reason, the terms containing odd powers of $S$
are neglected. As a result,
\begin{equation}
\hat{U}=-\frac{2\pi \hbar^2 }{m} \rho \left(
d+d_{1}S_{z}^{2}\right)  \label{N_hat}
\end{equation}
Suppose $m$ denotes magnetic quantum number, then for a
particle in a state that is an eigenstate of the spin projection operator $S_{z}$, 
one can express $U$ as
\begin{equation}
U\left( m\right) =-\frac{2\pi \hbar^2 }{m} \rho 
\left(d+d_{1}m^{2}\right)  \label{N_m}
\end{equation}

According to eq. (\ref{N_m}), the particle states with quantum numbers $m$
and $-m$ have the same $U$. 
As we see the picture of deuteron energy levels splitting in a medium,
coincides with that of splitting of atom energy levels in electric field due to Stark effect.
Thus we can guess that a deuteron in a medium undergoes the action of some effective
quasielectric nuclear field, caused by nuclear interactions of deuteron with proton (nucleus).

The spin wave function of deuteron captured to the trap at the time moment $T=0$ 
can be represented as a superposition of
basis spin wave functions $\chi _{m}$, which are eigenfunctions of the
operators $\hat{S}^{2}$ and $\hat{S}_{z},$ $\hat{S}_{z}\chi
_{m}=m\chi _{m}$:
\begin{equation}
\psi =\sum_{m=\pm 1,0}a^{m}\chi _{m}.
\end{equation}

Wave function of the
particle at time moment $T$ can be expressed as:
\begin{equation}
\Psi =\left\{ 
\begin{array}{c}
a^{1} \\ 
a^{0} \\ 
a^{-1}
\end{array}
\right\} =\left\{ 
\begin{array}{c}
a\,e^{i\delta _{1}}e^{-\frac{i}{\hbar}U_{1}T} \\ 
b\,e^{i\delta _{0}}e^{-\frac{i}{\hbar}U_{0}T} \\ 
c\,e^{i\delta _{-1}}e^{-\frac{i}{\hbar}U_{-1}T}
\end{array}
\right\} =\left\{ 
\begin{array}{c}
a\,e^{i\delta _{1}}e^{-\frac{i}{\hbar}U_{1}T} \\ 
b\,e^{i\delta _{0}}e^{-\frac{i}{\hbar}U_{0}T} \\ 
c\,e^{i\delta _{-1}}e^{-\frac{i}{\hbar}U_{1}T}
\end{array}
\right\}
\end{equation}
\noindent It should be reminded that $U_{1}=U_{-1}$

Let us choose coordinate system in which plane $\left( xz\right) $
coincides with that formed by vector $\langle \overrightarrow{S}\rangle $
(\bigskip $<\vec{S}\mathbf{>=}\frac{<{\psi }\mathbf{|}\vec{S}\mathbf{|}{\psi 
}>}{\mid \psi \mid ^{2}})$\ in time moment $T=0$. 
In this case $\delta _{1}-\delta _{0}=\delta _{-1}-\delta _{0}=0$
and components of vector at $T=0$ $\,\,\,\langle S_{x}\rangle \neq 0,\langle
S_{y}\rangle =0.$

As a result we obtain:
\begin{eqnarray}
&<&{S}_{x}>=\sqrt{2}
e^{-\frac{1}{2}\rho (\sigma _{0}+\sigma _{1})cT}
b(a+c)
\cos [\frac{2\pi \hbar \rho }{m}{Re}d_{1}\,T]
/|\psi |^{2},  \nonumber \\
&<&{S}_{y}>=-\sqrt{2}
e^{-\frac{1}{2}\rho (\sigma _{0}+\sigma _{1})cT}
b(a-c)
\sin [\frac{2\pi \hbar \rho }{m}{Re}d_{1}T]
/|\psi |^{2},  \label{S_}
\\
&<&{S}_{z}>=
e^{-\rho \sigma _{1}cT\,}
(a^{2}-c^{2})
/|\psi |^{2}, 
\nonumber
\end{eqnarray}

Particle with spin 1 also possesses tensor polarization i.e. tensor of
rank two {\ $\hat{Q}_{ij}=3/2(\hat{S}_{i}\hat{S}_{j}+\hat{S}%
_{j}\hat{S}_{i}-4/3\delta _{ij})$ } for it we can obtain
\begin{eqnarray}
&<&{Q}_{xx}>=
\left\{ 
-\left[ a^{2}+c^{2}\right] 
\frac{1}{2}\,e^{-\rho \sigma_{1}\,cT}+
{b}^{2}\,e^{-\rho \sigma _{0}\,cT}+
3e^{-\rho \sigma_{1}\,cT}\,ac\,
\cos \left[ \delta _{1}-\delta _{-1}\right] 
\right\} 
/|\psi|^{2}  \nonumber \\
&<&{Q}_{yy}>=
\left\{ -\left[ a^{2}+c^{2}\right] \frac{1}{2}\,
e^{-\rho \sigma_{1}\,cT}+
{b}^{2}\,e^{-\rho \sigma _{0}\,cT}-
3e^{-\rho \sigma_{1}\,cT}\,ac\,\cos \left[ \delta _{1}-\delta _{-1}\right] \right\} /|\psi
|^{2}  \nonumber \\
&<&{Q}_{zz}>=
\left\{ \left[ a^{2}+c^{2}\right] \frac{1}{2}\,
e^{-\rho \sigma_{1}\,cT}-
2{b}^{2}\,e^{-\rho \sigma _{0}\,cT}\right\} /|\psi |^{2}~,
\label{quadr} \\
&<&{Q}_{xy}>=-3e^{-\rho \sigma _{1}\,cT}\,ac\,\sin \left[ \delta _{1}-\delta
_{-1}\right] /|\psi |^{2},  \nonumber \\
&<&{Q}_{xz}>=\frac{3}{\sqrt{2}}
e^{-\frac{1}{2}\rho (\sigma _{0}+\sigma_{1})cT}
b(a-c)
\cos [\frac{2 \pi \hbar \rho }{m}{Re}d_{1}T]
/|\psi|^{2},  \nonumber \\
&<&{Q}_{yz}>=-~\frac{3}{\sqrt{2}}
e^{-\frac{1}{2}\rho (\sigma _{0}+\sigma_{1})cT}
b(a+c)
\sin [\frac{2 \pi \hbar \rho }{m}{Re}d_{1}T]
/|\psi|^{2}~,  \nonumber
\end{eqnarray}
where $|\psi |^{2}=2{}(a^{2}+c^{2})\,
e^{-\rho \sigma_{1}\,cT}+{}b^{2}\,
e^{-\rho \sigma _{0}\,cT}$.

According to (\ref{S_},\ref{quadr}) rotation appears if the angle
between polarization vector and momentum of a particle differs from $\frac{\pi 
}{2}$. At this for acute angle between polarization vector and momentum the
sign of rotation is opposite than that for obtuse angles.

If spin is orthogonal to momentum then $(a=c)$ particle spin (tensor of
quadrupolarization) oscillate (do not rotate).

Let us evaluate phase of oscillation
\begin{equation}
\varphi =\frac{2\pi \hbar \rho \;T}{m}{Re}d_{1}.
\label{fi}
\end{equation}
Density of particle beam passing through a trap can be estimated from the
following discourse. Suppose $N_c$ is the number of particles in storage ring,
they are distributed over the volume $V=SL$, where $S$ is the cross section
of a beam rotating in storage ring and $L$ is the length of closed orbit.
Thus, ${\rho}_c=\frac{N_c}{SL}$. For rotation frequency $\nu$ density of particle
increases $\nu$ times $\rho={\rho}_c \nu=\frac{N_c \nu}{SL}$.

For $N_c={10}^{11}$, $\nu=1.6 \cdot 10^6$ 
(for example, for COSY),
$S=1$ $cm^2$, $L \cong 2 \cdot 10^4 cm$ it is easy to calculate that $\rho \simeq 10^{13}$.
Considering that for deuteron scattering by hydrogen
$Re d_1 \sim 10^{-13}$ $cm$ phase of oscillation (\ref{fi}) 
can be written as $\varphi  \simeq 3 \cdot 10^{-3} \,T$.
One can estimate that for $T=10^3 sec$ the angle $\varphi=3$ $rad$.

So, the effect is considerable.

Attention should be drawn to the fact that hydrogen atom in triplet state has spin 1.
Therefore,
all the above is relevant for a cell containing hydrogen atoms in triplet state, too.
According to \cite{6rot} study of deuteron spin-dependent scattering amplitude
allows, among other things, to investigate real part of nucleon-nucleon scattering amplitude. 
If a cell contains hydrogen atoms, then studying spin oscillation and rotation of hydrogen atom
blowing by a flow of nonpolarized protons one can reconstruct real part of proton-proton scattering
amplitude.


\end{document}